\documentclass[fleqn,10pt]{wlscirep}
\usepackage[utf8]{inputenc}
\usepackage[T1]{fontenc}
\usepackage{subcaption,graphicx,color}
\usepackage{comment}
\usepackage{soul}
\usepackage{xcolor}
\title{Ultra-broadband photonic sampling of optical waveforms}

\author[1,2,*]{Dmitry A. Zimin}
\author[1,2]{Vladislav S. Yakovlev}
\author[1,3]{Nicholas Karpowicz}
\affil[1]{Max-Planck-Institut für Quantenoptik, Hans-Kopfermann-Strasse 1, 85748 Garching, Germany}
\affil[2]{Fakultät für Physik, Ludwig-Maximilians-Universität, Am Coulombwall 1, 85748 Garching, Germany}
\affil[3]{CNR NANOTEC Institute of Nanotechnology, via Monteroni, 73100 Lecce, Italy}

\affil[*]{d.a.zimin@gmail.com}

\begin{document}

\flushbottom
\maketitle

\noindent\textbf{Direct access to the electric field of light\cite{Kienberger2004} provides a subcycle view of the polarization response of matter\cite{sommer2016}, thus enabling sensitive metrology in physics\cite{Hohenleutner2015, Kuehn2010, Gaal2007, Ulbricht2011, Schlaepfer2018}, chemistry\cite{Mukamel2016}, and medicine\cite{Huber2021, Pupeza2020}.
Optical-field sampling techniques take advantage of some process that lasts much less than an optical cycle of the measured light wave.
Most of them rely on the generation of free electrons either by a weak extreme ultraviolet pulse\cite{Hentschel2001,Itatani2002, Goulielmakis1267} or by a strong optical pulse\cite{Thunich2011, Schiffrin2012, Park2018, Kubullek2020, Sederberg2020, Zimin2021}.
The ability of such techniques to detect weak signals is limited by undesirable effects associated with ionization. 
An alternative is all-optical methods, where the fast process is a nonlinear wave mixing.
Such photonic methods may rely on extreme ultraviolet\cite{Kim2013, Wyatt2016, Carpeggiani2017} or optical pulses \cite{Leitenstorfer1999, Keiber2016,Karpowicz2008, Dai2006}.
They typically have better sensitivity\cite{Leitenstorfer1999, Riek420, Pupeza2020}, but smaller bandwidth.
Here we propose generalized heterodyne optical-sampling techniques (GHOSTs), which overcome the bandwidth limit through controlling light-pulse waveforms.
} 

\begin{figure}[ht!]
\centering
	\includegraphics[scale=0.55]{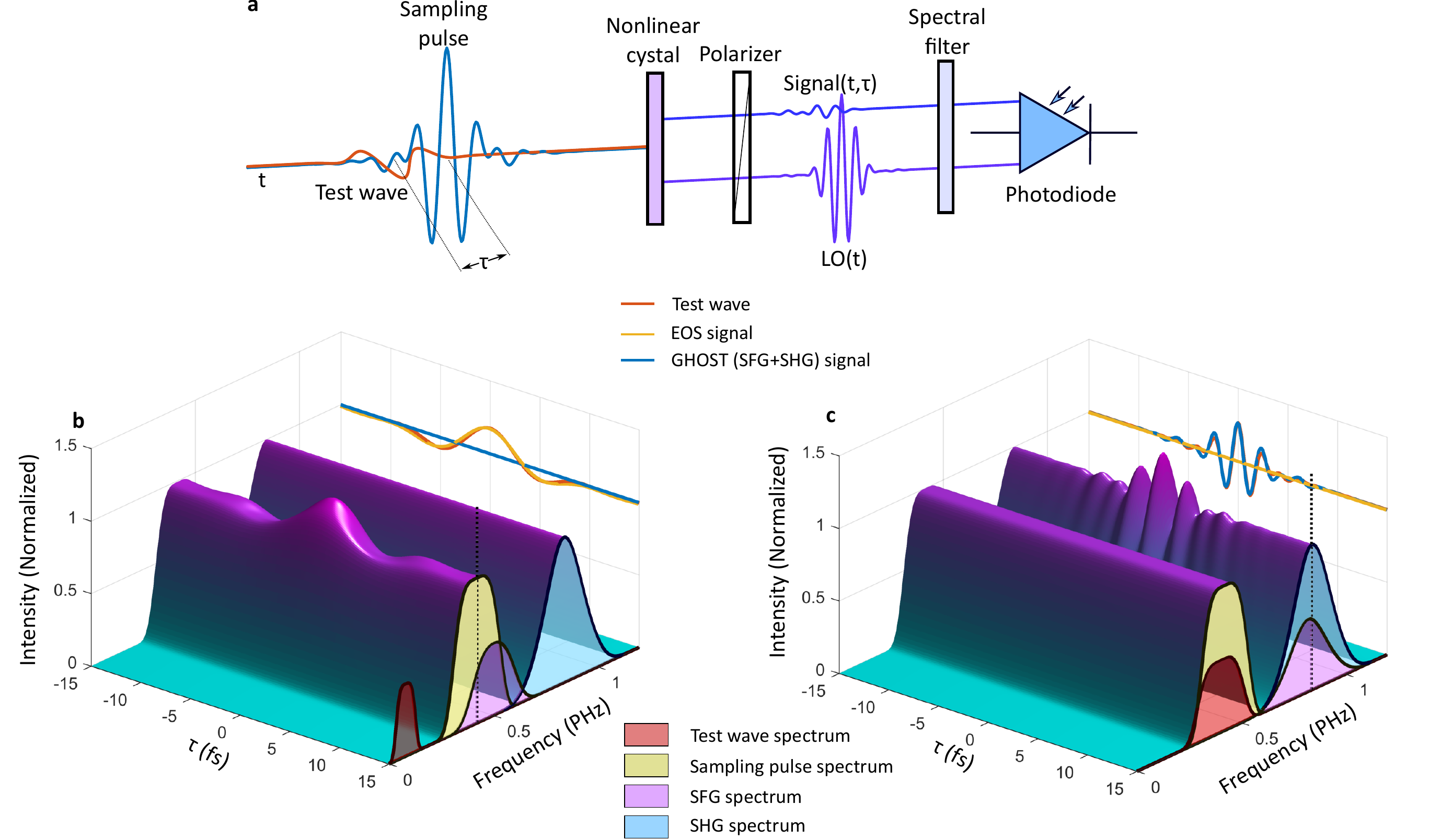}
	\caption{\textbf{Illustration of the measurement concept.}
	\textbf{a}, Two pulses are incident on the nonlinear medium: a test waveform that we wish to measure and a sampling pulse that will be used to probe its structure.
	Two waves emerge from the crystal: a local oscillator (LO), which is formed by nonlinear propagation of the sampling pulse, and a signal, which is the product of nonlinearly mixing both input pulses.
	These pass through a polarizer and a bandpass spectral filter, and the intensity of the light is recorded in a photodiode.
	\textbf{b}, In the case of electro-optic sampling, sum-frequency generation forms the signal that primarily overlaps with the unmodified spectrum of the sampling pulse.
	The interference of the signal with the local oscillator modulates the spectral intensity vs.\ time delay $\tau$. At the detection frequency, labeled with a vertical dashed line, the amplitude traces out the electric field waveform, providing the EOS signal.
	\textbf{c}, For a test pulse in the visible spectral range, the SFG spectrum lies at higher frequencies, predominantly in the ultraviolet, and possesses little spectral overlap with the sampling pulse.
	However, it significantly overlaps with the second harmonic of the sampling pulse that was generated in the same nonlinear crystal, permitting the SFG+SHG GHOST to take place.
	The modulation of spectral intensity in the UV band, at the detection frequency marked by the dashed line, traces out the test waveform.
	This concept can be extended to detection of light in other spectral regions, requiring a combination of nonlinear processes to produce spectrally overlapping signal and LO waves.}
	\label{fig:eos_ghost_concept}
\end{figure}

Let us review the fundamental form of nonlinear-optics-based field-sampling techniques that exist for the infrared and terahertz domains (see Fig.~\ref{fig:eos_ghost_concept}).
In these techniques, the field that is being sampled (test field) undergoes nonlinear wave mixing with a sampling pulse that creates a temporal gate.
The outcome of this nonlinear interaction is referred to as heterodyne signal or, for brevity, simply signal, even though this is not yet the outcome of the measurement.
This signal is first made to interfere with the local oscillator (LO), which is what the sampling pulse becomes after transmission through the same nonlinear medium in the absence of the test field.
In a spectral region where the signal and the LO overlap, the spectral intensity of their superposition depends on the delay between the test and sampling pulses.
Two prominent examples of such measurement techniques are electro-optic sampling (EOS)\cite{Leitenstorfer1999, Keiber2016} and air-biased-coherent-detection (ABCD)\cite{Karpowicz2008, Dai2006}.
In EOS, the signal is produced from sum and difference frequency generation between the test waveform and sampling pulse, and the LO is provided by the unperturbed sampling pulse \cite{Porer2014, Keiber2016}.
In ABCD, the signal is the result of four-wave mixing, summing two photons from the sampling pulse and subtracting one from the test waveform, while the LO is provided by field-induced second harmonic generation: a bias applied to the nonlinear medium allows it to combine two sampling-pulse photons even if the medium is centrosymmetric. The phase of the nonlinearly-created local oscillator is modulated by changing the direction of the bias field. This modulation creates sidebands of the LO which interfere with the signal, thus enabling heterodyne detection of pulses over a continuous range from the THz through the mid infrared.\cite{Karpowicz2008}.

In addition to the spectral overlap between the signal and the LO, these schemes had to satisfy two key requirements: \textbf{1}: only one photon from the test waveform is involved in the generation of the signal, and \textbf{2}: the same number of photons from the sampling pulse is involved in forming the LO and signal.
The former one is necessary for linear detection of the field.
The latter one ensures that the interference between the signal and the LO is insensitive to the carrier-envelope phase (CEP) of the sampling pulse, which is known as the global phase\cite{Hassan2016} for arbitrary waveforms.
If requirement \textbf{2} is satisfied, the phase cancels during heterodyne detection, meaning that sampling pulses with unstable CEP can be used.
When both these requirements are fulfilled, the interference between the signal and local oscillator has an amplitude and phase determined by the test waveform, and varying the delay between the pulses while recording the resulting intensity measures the electric field of the test waveform.

Requirement \textbf{2} is responsible for the primary limit on the maximum detection frequency of the techniques: since the signal and LO wavelengths must match, and the signal frequency is shifted by the frequency of the test waveform, it cannot exceed the bandwidth of the local oscillator.
However, the emergence of CEP-stabilized mode-locked laser oscillators allows us to bypass this restriction.
For example, using second harmonic generation to form the LO for a signal based on sum-frequency generation, i.\,e., adding one more photon from the sampling pulse to the LO of EOS, increases the frequency of the detection band by the carrier frequency of the sampling pulse.
We label this GHOST as SFG+SHG.
A $<5$-fs sampling pulse at 400 THz carrier frequency and 200-THz bandwidth, capable of detecting 0-200 THz via EOS, could in principle detect 100-700 THz via SFG+SHG (Fig.~\ref{fig:eos_ghost_concept}) or 900-1500 THz via DFG+SHG. Other GHOSTs, on the other hand, may in principle be used to detect even higher spectral components. Employing CEP-stable pulses, the detection bandwidth and spectral ranges can be expanded and tailored by choosing appropriate combinations of GHOSTs, and detection frequencies (see SI).
The price that one pays for this is that the phases of the LO and signal experience a different phase shift upon a change of the CEP of the sampling pulse.
As a result, a phase shift $\Delta\phi_\mathrm{CE} = a\left(N_\mathrm{S} - N_\mathrm{LO}\right)\phi_\mathrm{S}$ is applied to the measured waveform, where $N_\mathrm{S}$ is the number of sampling pulse photons summed over to arrive at the signal, $N_\mathrm{LO}$ is the number summed over to obtain the LO, and $a$ is 1 or -1 depending on the mixing process, and $\phi_\mathrm{S}$ is the CEP of the sampling pulse (for further details, see SI). 

In both EOS and ABCD, $\Delta\phi_\mathrm{CE} = 0$, while in the SFG+SHG GHOST, $\Delta\phi_\mathrm{CE} = -\phi_\mathrm{S}$.
This has a straightforward interpretation in the time domain: in any linear electric-field measurement, the detected waveform is the convolution of the true electric field with the response function of the detection system.
The addition of an unbalanced photon in the detection scheme causes this response function to oscillate in time, which allows for the detection of more rapidly varying fields, but repeated measurements of fields with reproducible (CEP-stabilized) waveforms will average to zero if $\phi_\mathrm{s}$ is not stabilized.

\begin{figure}[ht!]
\centering
	\includegraphics[scale=1.0]{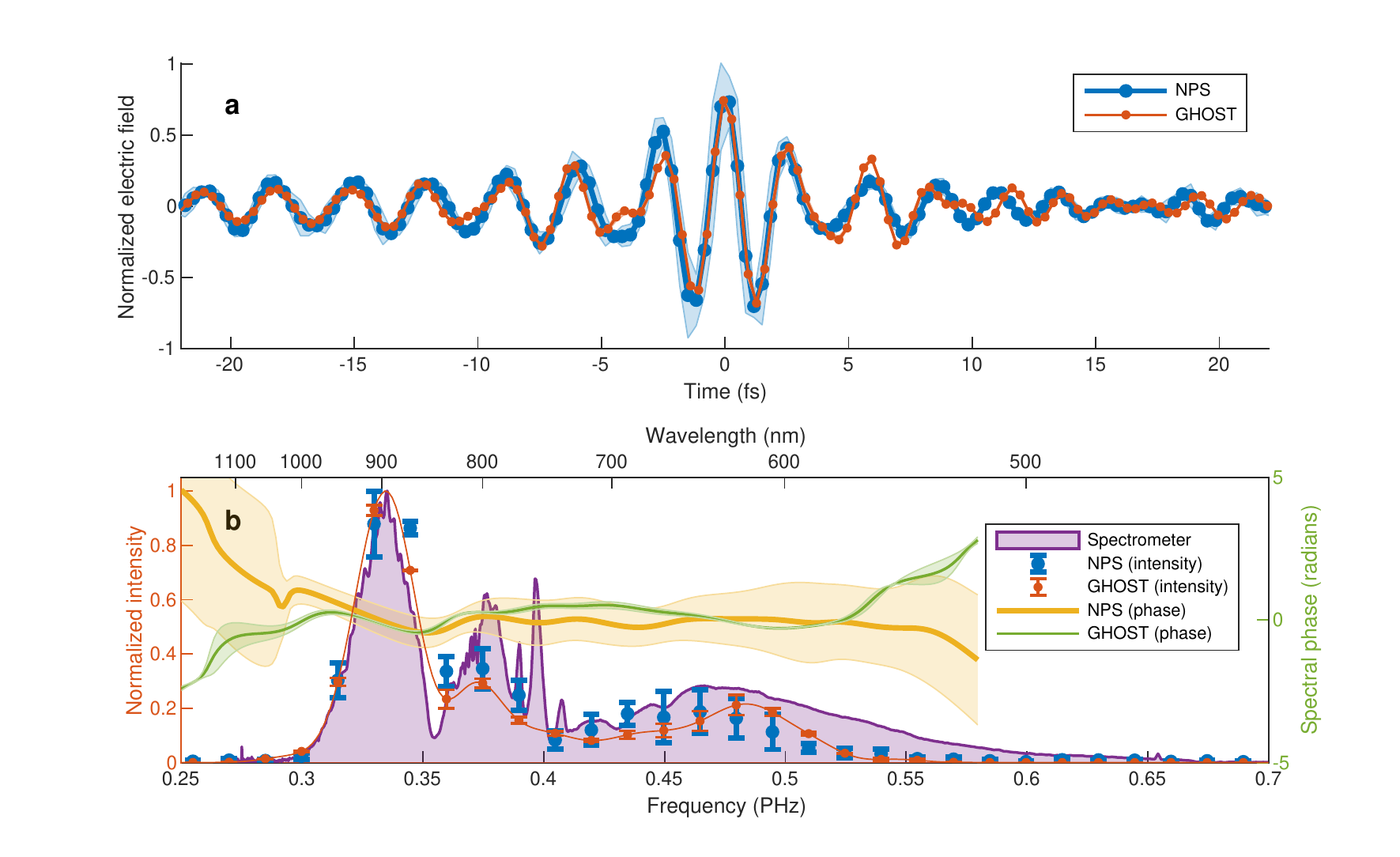}
	\caption{\textbf{Validation of GHOST detection via benchmarking against NPS and spectrometry.}
	\textbf{a}, Electric-field waveforms measured with the SFG/DFG+THG GHOST and NPS.
	The shaded areas represent the standard deviations evaluated from 3 individual measurements.
	\textbf{b}, The spectral intensities and phases of the waveforms shown in panel (a), as well as the spectrum obtained by a grating spectrometer.
	The shaded areas show the standard deviations for the spectral phases.
	We depicted the standard deviations of the spectral intensities with error bars.}
	\label{fig:benchmarking}
\end{figure}

To confirm the validity of the concept, we experimentally investigated the ``SFG/DFG+THG'' GHOST, where sum-frequency generation (SFG) and difference-frequency generation (DFG) together form the heterodyne signal, while third-harmonic generation (THG) forms the local oscillator (see SI for further details).
As a nonlinear medium that possesses both $\chi^{(2)}$ and $\chi^{(3)}$ nonlinearities, we chose a thin quartz crystal (see SI for details) for its low absorption and dispersion.
For detection, we used a bandpass filter that transmitted light between 0.83~PHz and 0.86~PHz (see SI).
Even though this frequency range is in the middle of the second harmonic of the sampling pulse, the orientations of the crystal and the polarizer were chosen such that only the $\chi^{(3)}$ nonlinearity contributed to the local oscillator.
We benchmarked this implementation of GHOST against nonlinear photoconductive sampling (NPS)\cite{Sederberg2020}, which is a field-sampling technique where temporal gating is achieved through highly nonlinear photoinjection of charge carriers.
Both GHOST and NPS measurements were performed using a 2.7-fs sampling pulse\cite{Sederberg2020}, and we used theoretical response functions to retrieve time-dependent electric field from the raw measured data (see SI for details).
The benchmarking results are shown in Fig.~\ref{fig:benchmarking}.
In the time domain (Fig.~\ref{fig:benchmarking}a), the two waveforms have very similar shapes: the Pearson correlation coefficient between the waveforms is $\rho = 0.93$.
Note that GHOST measurements show much smaller stochastic fluctuations---the peak standard deviation is 10 times smaller than that in the NPS measurements.
Also in the frequency domain (Fig.~\ref{fig:benchmarking}b), we see that spectral intensities and spectral phases have much smaller fluctuation in the GHOST measurement, as compared to the NPS one.
The spectra of both electric-field waveforms agree well with the spectrum obtained with a calibrated grating spectrometer (Ocean Optics).

\begin{figure}[ht!]
\centering
	\includegraphics[scale=1.0]{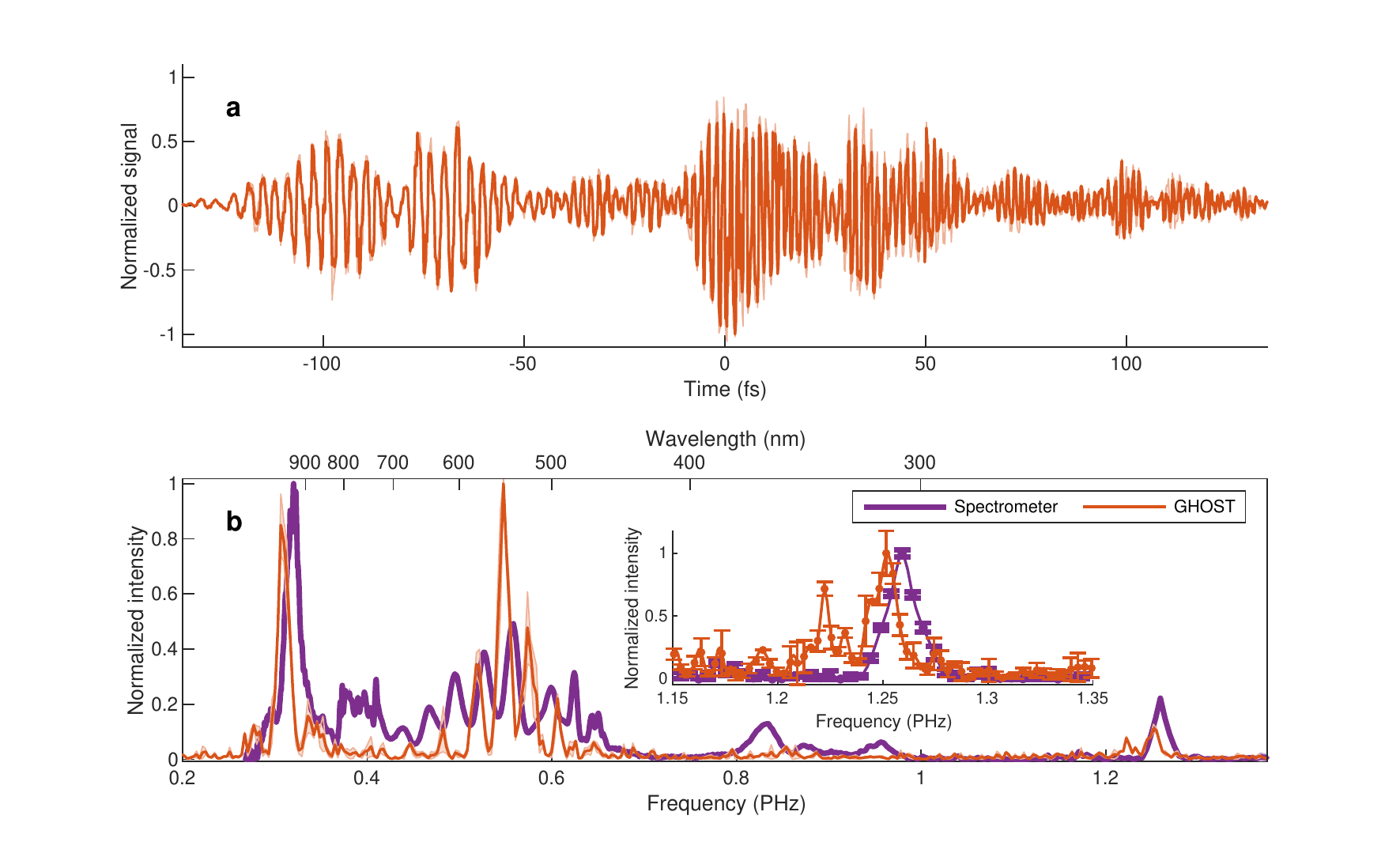}
	\caption{\textbf{Ultraviolet waveform detection with the DFG+THG GHOST.}
	We demonstrate the potential bandwidth of GHOST by applying it to a test waveform generated by up-converting the pulse shown in Fig.~\ref{fig:benchmarking}.
	The shaded areas and errorbars represent standard deviations.
	\textbf{a}, The detected waveform contains a rapid ultraviolet oscillation preceded by the lower-frequency light (detected through the SFG+THG GHOST).
	The delay between the two parts of the test pulse is caused by the group-velocity mismatch in the 100~{\textmu}m BBO crystal that we used for up-conversion.
	\textbf{b}, The spectrum of the GHOST waveform (red) and the spectrum measured by a grating spectrometer (purple).
	}
	\label{fig:ultraviolet}
\end{figure}

The benchmarking results shown in Fig.~\ref{fig:benchmarking} confirm only the validity of the SFG+THG GHOST --- to validate the DFG+THG GHOST, the spectrum of the test pulse must extend well above the frequencies that reach the photodiode.
As a simple estimation, let us take 0.3~PHz as the lowest frequency present in the sampling pulse and 0.8~PHz as the detection frequency.
In this case, we expect the spectral range covered by the DFG+THG GHOST to begin at 1.1~PHz.
So, we upconverted the test pulse in a 100~{\textmu}m BBO crystal, obtaining pulses with a spectrum that extended up to 1.25~PHz (see section 4 of SI).
We plot the GHOST waveform in Fig.~\ref{fig:ultraviolet}a (here, no attempt is made to account for the theoretical response function).
In Fig.~\ref{fig:ultraviolet}b, we compare the spectrum of the detected waveform with the spectrum that we measured with a grating spectrometer.
The inset displays the spectral range that is relevant to the DFG+THG GHOST.
While there are discrepancies between the two spectra, and a careful benchmarking in this spectral range is yet to be done, we conclude that our implementation of GHOST is capable of optical-field sampling at frequencies that exceed 1~PHz.

\begin{figure}[ht!]
\centering
	\includegraphics[scale=0.3]{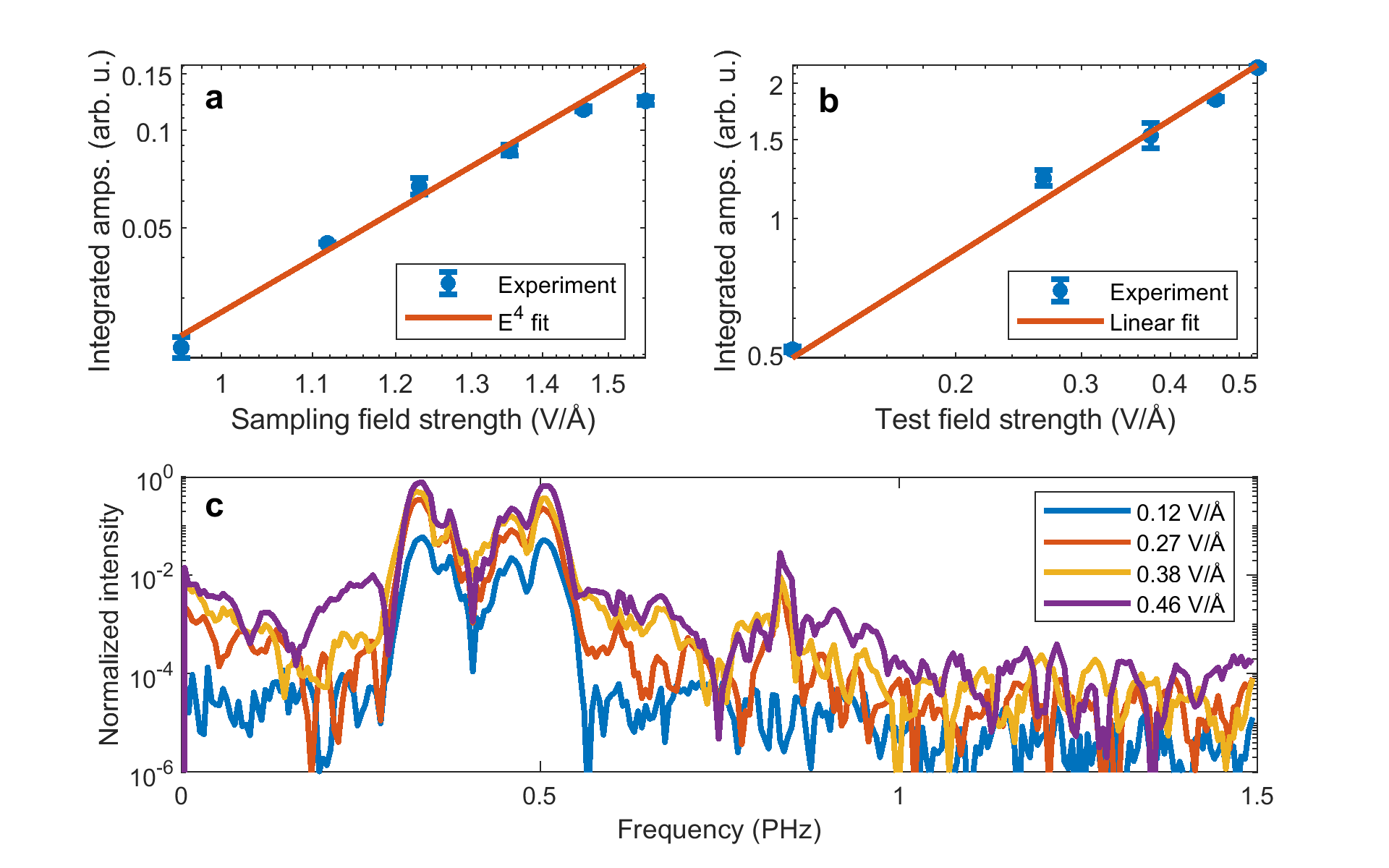}
	\caption{\textbf{Intensity scaling and signal-to-noise ratio.}
	(\textbf{a}) Scaling of the measured signal vs.\ field strength of the sampling pulse.
	(\textbf{b}) Scaling the measured signal vs.\ field strength of a test pulse.
	(\textbf{c}) Logarithmically-scaled typical measured spectrum and spectral noise floor. }
	\label{fig:scaling}
\end{figure}

As an additional verification that we indeed implemented the SFG/DFG+THG GHOST, we investigated how the measured waveforms depend on the peak strengths of the incident sampling and test pulses.
The results are show in Fig.~\ref{fig:scaling}.
As expected, the amplitudes of measured waveforms scale as the fourth power of the peak sampling field, $E_\mathrm{s}$.
Indeed, each photon in the heterodyne signal emerges with the involvement of one photon from the sampling pulse, while each photon in the local oscillator emerges from a three-photon nonlinear process.
Thus, the electric field of the heterodyne signal scales linearly with $E_\mathrm{s}$, the electric field of the local oscillator scales as $E_\mathrm{s}^3$, and the interference between the two fields makes the intensity oscillate with an amplitude that is proportional to $E_\mathrm{s}^4$.
We observe this dependence up to sampling fields as strong as 1.5~V/{\AA} (see Fig.~\ref{fig:scaling}a).
At this field strength, multiphoton photoinjection of charge carriers becomes significant.
According to Fig.~\ref{fig:scaling}b, the amplitude of the measured waveforms scales linearly up to test fields as strong as 0.5~V/{\AA}.
Nevertheless, Fig.~\ref{fig:scaling}c shows that already the 0.27-V/{\AA} test pulse produces a waveform that has high-frequency components that are absent in the waveform measured with the 0.12-V/{\AA} test pulse.
From Fig.~\ref{fig:scaling}c, we can also estimate a typical signal-to-noise ratio (SNR) of our measurements.
By comparing the noise baseline with the peak signal strength, we obtain an intensity SNR of approximately 30 dB independently of the test-field strength, which indicates that, in these measurements, the noise was dominated by the fluctuations of laser intensity.
Future improvements (e.g., using a balanced detection scheme) may further improve the SNR.

In summary, generalized heterodyne optical-sampling techniques present a promising approach for measuring electric fields of broadband laser pulses.
We studied one particular combination of nonlinear optical processes: ``SFG/DFG + THG'', where the $\chi^{(2)}$ nonlinearity mixed the sampling and test waves, thus generating a heterodyne signal, while the third harmonic of the sampling pulse provided a local oscillator.
This fidelity of this technique was much higher than that of nonlinear photoconductive sampling.
In principle, petahertz bandwidth is achievable with GHOST.
The spectral response of GHOST can be tailored for a particular application through the choice of nonlinear processes and the frequency range that reaches the photodetector.
Any medium that exhibits the required nonlinearities, such as solids or gases, could be used for waveform sampling.
Since they rely on low-order nonlinearities, GHOSTs can be used with a broad variety of laser systems that produce pulses with a stabilized carrier-envelope phase.

Direct time-domain measurements of light-matter interaction with sub-cycle resolution are at the forefront of attosecond science.
These new techniques for direct electric field measurement present new opportunities for highly sensitive time-resolved spectroscopy and for the extension of field-resolved metrology to new regimes of both wavelength and intensity. 
The flexibility offered in terms of signal photon energy, detection band, and nonlinear optical processes provides an opportunity for field measurements to be integrated into a variety of new investigations.

\section*{Acknowledgements}

This research is based upon work supported by the US Air Force Office of Scientific Research under award number FA9550-16-1-0073.
This work was partially supported by the Air Force Office of Scientific Research (MURI, grant FA9550-14-1-0389 and grant FA9550-16-1-0156), by the Munich Centre for Advanced Photonics, and by the IMPRS-APS.
N.K.\ was partially supported by the FISR-CNR project “TECNOMED—Tecnopolo di nanotecnologia e fotonica per la medicina di precision”.
The authors thank Matthew Weidman for helpful discussions and Ferenc Krausz for initiating the research that created the prerequisites for this study as well as for helpful discussions.

\section*{Author contributions statement}

D.Z. proposed the concept and performed the measurements with the support of N.K.
The data analysis was performed by D.Z., N.K. and V.S.Y.
The theoretical simulations were conducted by D.Z.\ and N.K with the support of V.S.Y.
The manuscript was written by N.K, D.Z and V.S.Y.

\section*{Methods}

The laser system used for the experiments (more details in section 1 of SI) comprises a Ti:sapphire oscillator (Rainbow 2, Spectra Physics), followed by chirped pulse amplification to 1 mJ pulse energy at 3 kHz repetition rate, further spectral broadening in a hollow-core fiber and a chirped mirror compressor. 

The experimental data acquisition (more details in section 2 of SI) was performed with a dual-phase lock-in amplifier (SR-830, Stanford Research Systems) as well as with a grating spectrometer (Ocean optics).

\bibliography{ms} 
\bibliographystyle{naturemag-doi}

\end{document}


\flushbottom

\section*{Supplementary information}

\subsection{Optical system}
\label{laser_beamline}

\begin{figure}[ht!]
\centering
	\includegraphics[scale=0.6]{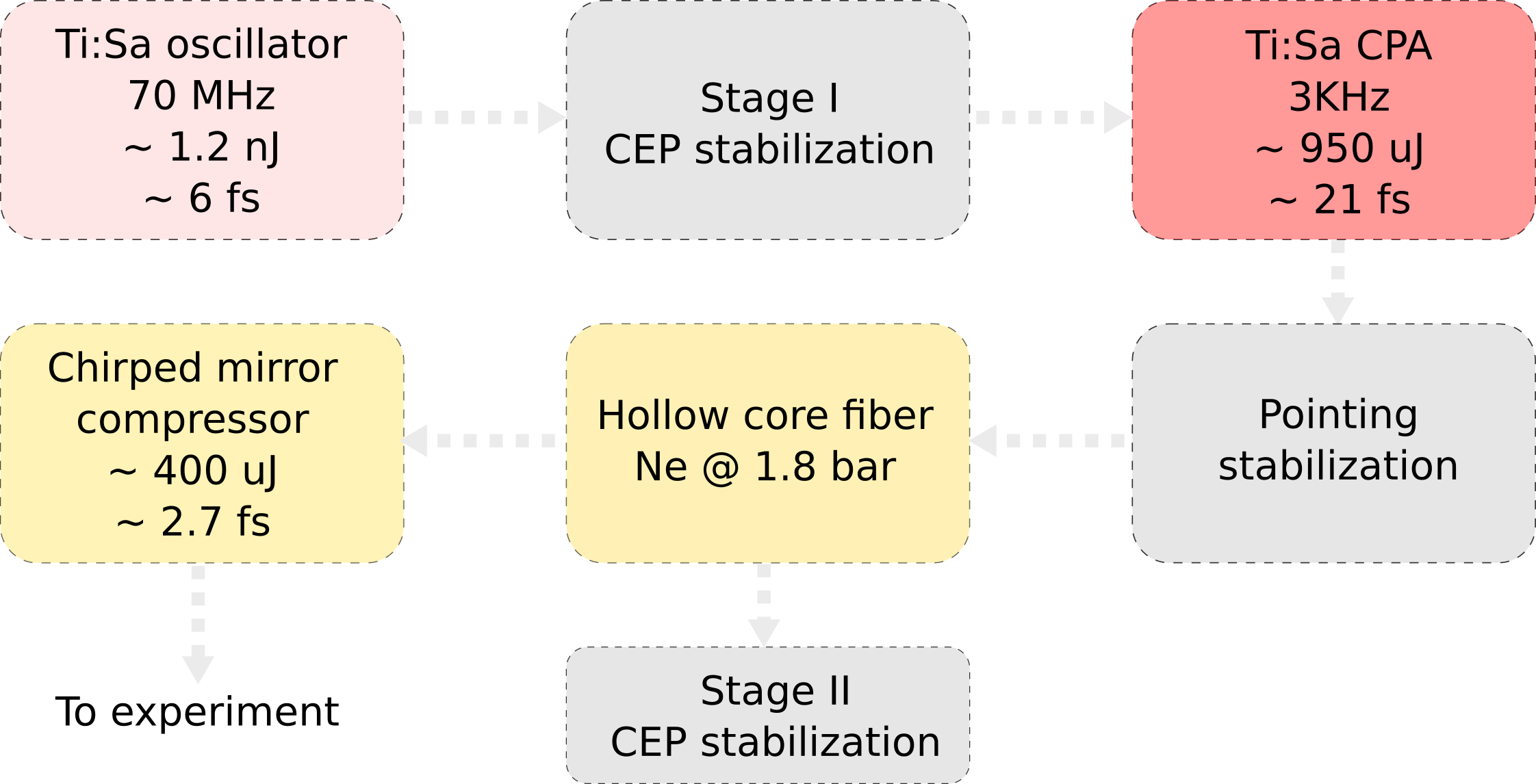}
	\caption{Diagram of the experimental beam-line}
	\label{fp2_scheme}
\end{figure}

Fig.~\ref{fp2_scheme} shows a schematic of the laser beamline\cite{Cavalieri_2007, Tim_thesis} used for the experiments.
Ti:sapphire oscillator (Rainbow 2, Spectra Physics) provides an octave-spanning bandwidth with about $\sim 750$~nm central wavelength.
The output from the oscillator is guided through the \textit{stage I} CEP stabilization module based on the feed-forward scheme\cite{Tim_thesis}.
CEP stable pulses are then further amplified within a 9-pass cryo-cooled Ti:sapphire chirped-pulse amplifier at a repetition rate of 3~kHz, and temporally compressed using a transmission grating based compressor yielding $\sim 21$~fs pulses with $\sim 2.5$~W output power.
The amplified pulses are further spectrally broadened in a hollow-core fiber (HCF), filled with 1.8-bar neon gas resulting in a spectral broadening to a bandwidth of about 400 - 1100~nm.
The chirped mirror compressor consisting of 3 pairs of chirped mirrors in combination with about 6~mm of fused silica glass and 50~cm of air is then used to compress the broadened spectrum to about $\sim 2.7$~fs (FWHM) duration pulse\cite{Sederberg2020}.
The pulse width is based on the intensity envelope of the complex-valued analytic wave, whose real part is the electric field measured by NPS, and the imaginary part is the Hilbert transform of the field.
The compressed pulses are then guided to an experimental setup for further experiments.
The small fraction of spectrally broadened light is reflected from the Brewster window of the hollow core fiber sealing and guided to a 'Stage II CEP stabilization'.
The Stage II stabilization is based on f-to-2f technique~\cite{schultze_thesis}.

\subsection{Data acquisition}
\label{aquasition}

\vskip 4mm

The signal from GHOST detection was a current generated by a photodiode (ALPHALAS).
This current was first converted to voltage with further amplification by a transimpedance amplifier (DLPCA-200, FEMTO Messtechnik).
The amplified voltage signal was then connected to a dual-phase lock-in amplifier (SR-830, Stanford Research Systems), triggered by an electrical signal synchronized with a half of the repetition rate of the laser.

The measured signal from the lock-in amplifier was read by software on a computer via GPIB interface (National Instruments).

\subsection{Sample medium characterization}
\label{thickness_section}

\vskip 4mm

In order to precisely determine a spectral response of the GHOST optical sampling techniques, the propagation effects within a sampling medium should be considered.
For this, the precise thickness of the sampling medium must be known.
The thickness of a solid medium can be determined with a conventional spectrometer and a broadband light source.
An optical pulse gets reflected at both surfaces of a sample.
After an even number of reflections, light travels in the same direction as the incident pulse, but it is delayed due to a longer propagation through the medium.
Multiple pulses in the time domain will result in spectral fringes, which can be measured with a spectrometer.
The total transmission of a medium, which takes into account reflections at the surfaces, is given by
\begin{equation}
T(n,\omega, d) = \frac{1}{\cos \left( \frac{\omega}{c}n d \right) - \frac{i}{2} \left( n + \frac{1}{n} \right) \sin \left( \frac{\omega}{c} n d \right)},
\label{thickness_equation}
\end{equation}
where $T$ is an intensity transmission, $n$ is a real valued refractive index of a frequency $\omega$, $c$ is the vacuum speed of light, and $d$ is a sample thickness.

\begin{figure}[ht!]
\centering
	\includegraphics[scale=0.3]{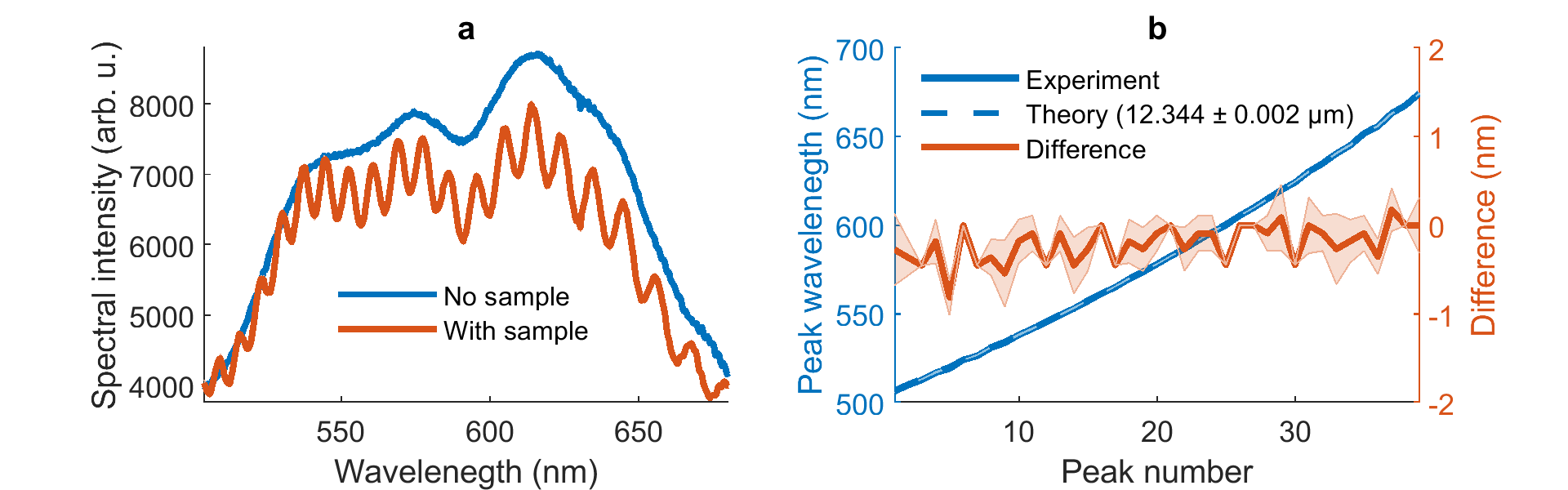}
	\caption{Measured change of spectral intensities  with a $z$-cut $\alpha$-quartz sample (\textbf{a}).
	Positions of spectral fringes allow to extract a sample thickness for the sample (\textbf{b}).}
	\label{thickness_measurement}
\end{figure}

To precisely measure the thickness of $z$-cut $\alpha$-quartz sample, the intensity spectrum of a broadband pulse (section \ref{laser_beamline}) was measured with and without the sample (Fig.~\ref{thickness_measurement}).
The spacing of fringes was then modeled with Eq.~\eqref{thickness_equation}, where we evaluated the frequency-dependent refractive index using the Sellmeier equation\cite{Ghosh1999}.
The calculated sample thickness is the one that provides the best agreement of fringe positions in the experiment.

The $z$-cut $\alpha$-quartz crystal sample was provided by MTI Corporation, and its thickness was further reduced by polishing.

\subsection{Optical schematics of experimental setups}
\label{schematics}

\vskip 4mm

\begin{figure}[ht!]
\centering
	\includegraphics[scale=0.1]{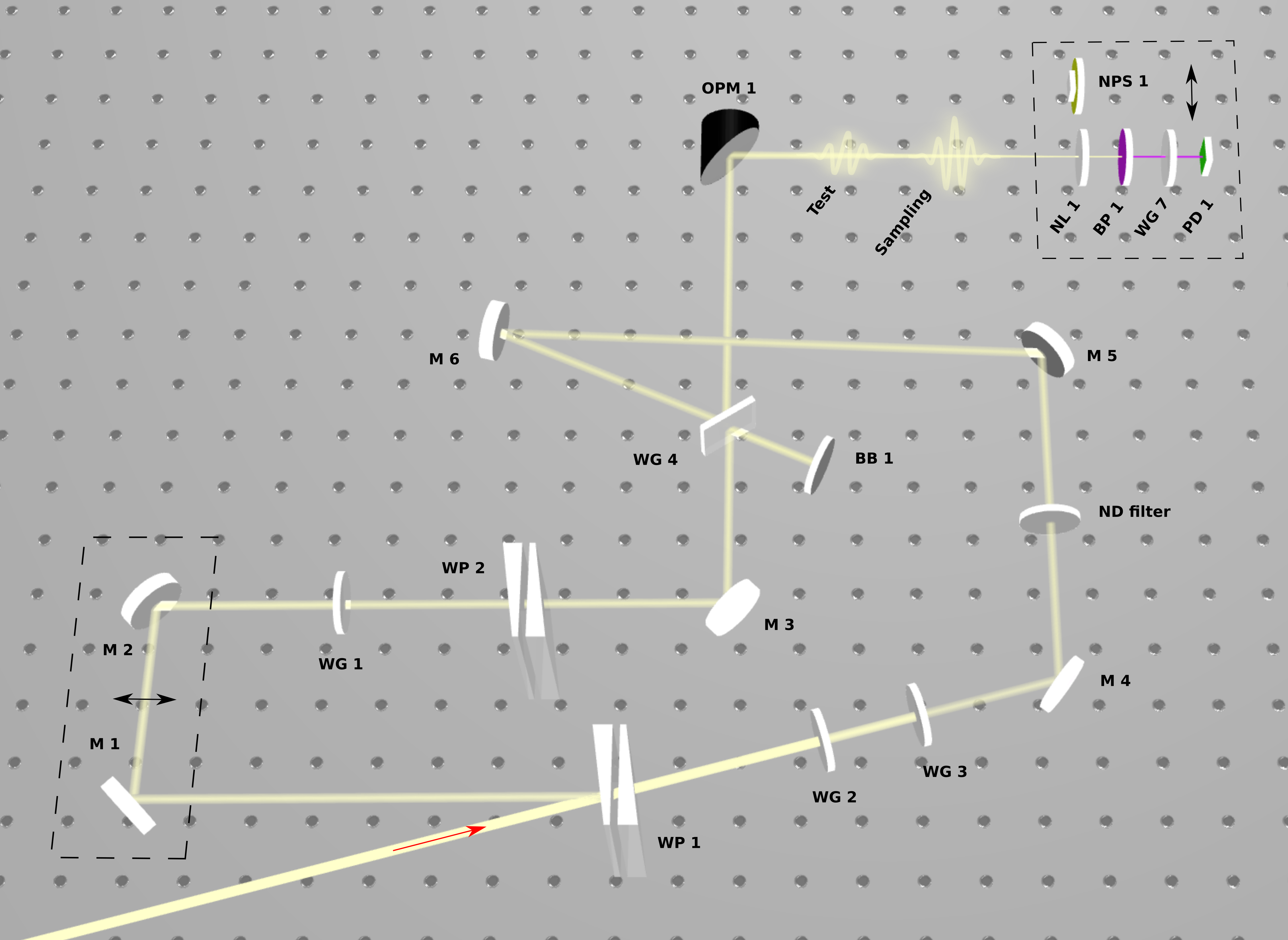}
	\caption{Optical setup for GHOST detection and characterization, as well as for the GHOST-NPS benchmarking in the visible spectral range.}
	\label{ghost_visible}
\end{figure}

The input laser beam is split into two arms by taking a reflection from the front surface of the first wedge in the wedge pair WP~1.
The reflected beam (sampling arm) is guided through a delay stage consisting of a retro-reflector M~1 and M~2 protected silver mirrors mounted on a closed-loop piezo stage (PX 200, Piezosystems Jena).

The wire-grid polarizers (WG~1 - WG~3) were used to attenuate the test and sampling pulses, while fused-silica wedge pairs (WP~1, WP~2) allowed us to compress and fine-tune the CEPs of the pulses.
Both arms were then recombined by a wire-grid polarizer WG~4, which is set to transmit the pulses in the test arm and reflect the pulses in the sampling arm.
The orthogonally polarized pulses were then focused by a protected silver off-axis parabolic mirror OPM~1 for the NPS or GHOST detection.
The quartz crystal ($\sim 12$~{\textmu}m thick, $z$-cut) NL~2 is used to generate a local oscillator and a heterodyne signal, which are then filtered by a bandpass filter BP~1 and the wire-grid polarizer WG~5.
The light is then detected by a photodiode.

The beam block BB~1 was used to block the residual light after the WG~4 polarizer, while plane protected silver mirrors M~3 - M~6 were used for beam guiding.

The reflective UVFS natural-density filter was used for field attenuation in the test arm.

\begin{figure}[ht!]
\centering
	\includegraphics[scale=0.1]{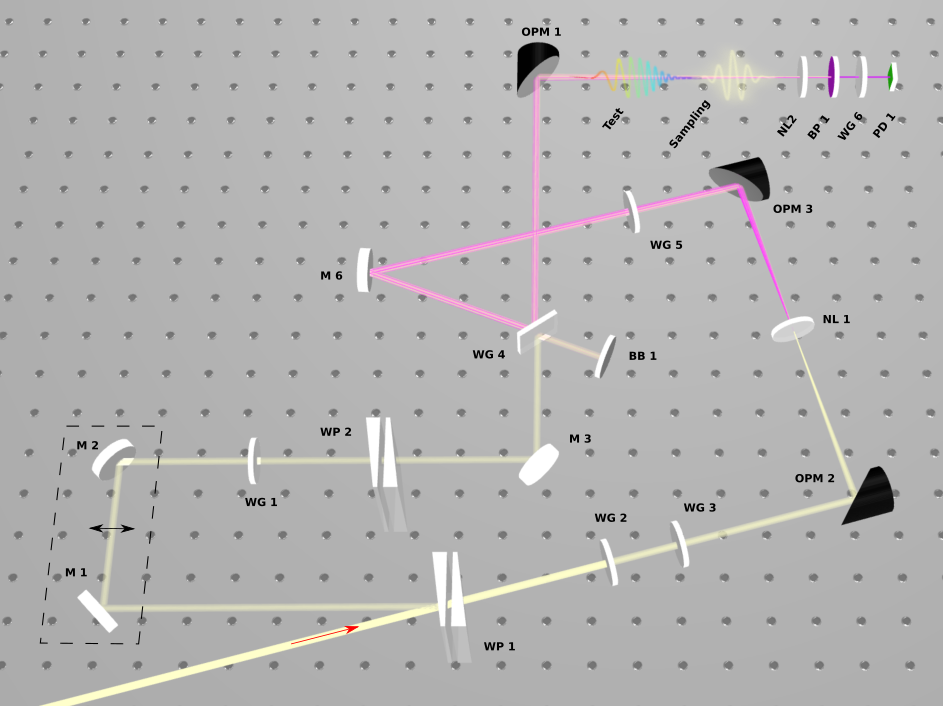}
	\caption{Optical setup for the broadband SFG+THG and DFG+THG GHOSTs.}
	\label{ghost_broadband_setup}
\end{figure}

The broadband GHOST detection was performed with a setup presented in Fig.~\ref{ghost_broadband_setup} as described below.

The protected aluminum off-axis parabolic mirror OPM~2 focused the beam onto a BBO crystal (100~{\textmu}m thick, type II, $\theta = 29.2^\circ$) NL~1 to generate new spectral components down to 235~nm.
The fundamental light with the newly generated spectrum is re-collimated by a protected aluminum off-axis parabolic mirror OPM~3. 

Both arms were then recombined in a wire-grid polarizer WG~4 and focused with a protected aluminum off-axis parabolic mirror to the GHOST detection setup.

A plane-protected silver mirror M~3 and a plane-protected aluminum mirror M~6 were guided the beam.

The wire-grid polarizer WG~5 ensured the linear polarization of the test arm after interaction with a BBO crystal NL~1.

\subsection{Maxwell-equation model with propagation effects}
\label{propagation_model}

\vskip 4mm

Propagation effects, such as dispersion, absorption, phase mismatch, etc., influence the spectral response of the GHOST detection.
We modeled these effects by numerically solving Maxwell's equations on a one-dimensional spatial grid.

Since the dispersion of the refractive index is a deciding factor in the coherent build-up of the signals of interest, it is important to approximate it well.
This is somewhat atypical for a time-domain calculation, and naive application of the refractive index from a Sellmeier equation found in the literature may lead to non-physical results, as they are typically the summation of a number of Lorentzian resonances, matched only to the real part of the refractive index.
However, in the time domain, the refractive index of the material is contained in a response function, which must obey causality.
Single Lorentzian resonances, although their resulting polarizations can be calculated quite efficiently in the time domain, typically lead to large inaccuracies in the absorption beneath the band-gap of typical solids.

In our model, an appropriate time-domain response function is found by first placing a series of absorption bands in the frequency domain and then calculating the resulting refractive index through the Kramers-Kronig relations, which guarantees that the resulting function will obey causality.
The refractive index is matched to literature values in a nonlinear least-squares fitting.
There is some freedom in choosing the length of the response function, but typically it contains 3000-15000 time points, with a time step of 6 attoseconds (0.25 atomic units).
This means that the history of the electric field must be stored at each point on the grid which contains the crystal.
In other words, previous values of the electric fields must be stored in order to calculate the new ones.
Additionally, the histories of the linear and nonlinear polarizations are stored at each point.

The linear polarization is calculated as a convolution of the field history with the response function, meaning its value at time $t$ is given by

\begin{equation}
P^{(1)}(t) = \int_{-t_L}^t dt'\,\chi^{(1)}(t)E(t-t'),
\end{equation}
where $t_L$ is the (finite) length of the response function $\chi^{(1)}(t)$.

The nonlinear polarizations can be calculated from the convolution of the field with a separate set of response functions, which can differ for different orientations of each instance of the field.

For the second-order polarization:
\begin{equation}
P_k^{(2)}(t) = \int_{-t_L}^t dt'\, \chi^{(2)}_{jk}(t)E_j(t-t') \int_{-t_L}^{t'} dt''\, \chi^{(2)}_{ij}(t)E_i(t-t'').
\end{equation}

For the third order polarization:
\begin{equation}
P_\ell^{(3)}(t) = \int_{-t_L}^t dt'\,\chi^{(3)}_{k\ell}(t)E_k(t-t') \int_{-t_L}^{t'} dt''\, \chi^{(3)}_{jk}(t)E_j(t-t') \int_{-t_L}^{t''} dt'''\, \chi^{(3)}_{ij}(t)E_i(t-t''')
\end{equation}

The subscripts of the fields and polarizations indicate the direction of the field within the material and, in general, must be summed over all combinations.
As a significant simplification, we employ the same response function, derived from the linear response, for the nonlinear terms, with a weighting determined by the susceptibility tensors of the crystal, thus approximating the dispersion of the nonlinear coefficients by Miller's rule. 

The values of the polarization in the given directions come from summing over the elements of the nonlinear tensors, which are then normalized by the known linear susceptibility $\chi^{(1)}$ at a frequency of interest $\omega_{0}$.
The normalization factor, $\chi^{(1)}$, can be extracted from a measured or calculated refractive index by means of Eq.~\eqref{ref_susc}.

\begin{equation}
n(\omega_0) = \sqrt{1 + \chi^{(1)}(\omega_0)}.
\label{ref_susc}
\end{equation}

The full polarization is calculated using the full tensor nature of the third and second-order nonlinearities, meaning that 14 separate polarizations are calculated for the two polarization components at each time step and contribute to the final polarization, which serves as a driving term in Maxwell's equations.
The calculation of the nonlinear polarization is by far the most numerically intensive step in the propagation.

The full propagation is performed using numerically evaluated spatial derivatives that are accurate to sixth order in the spatial grid step.
Convergence checks were performed such that the dispersion and transmission of broadband pulses up to 1 PHz frequency are transmitted with numerical dispersion accounting for $< 0.001$ of the actual dispersion.

\subsection{The spectral response of the photodiode and bandpass filter}

\vskip 4mm

\begin{figure}[ht!]
\centering
	\includegraphics[scale=0.32]{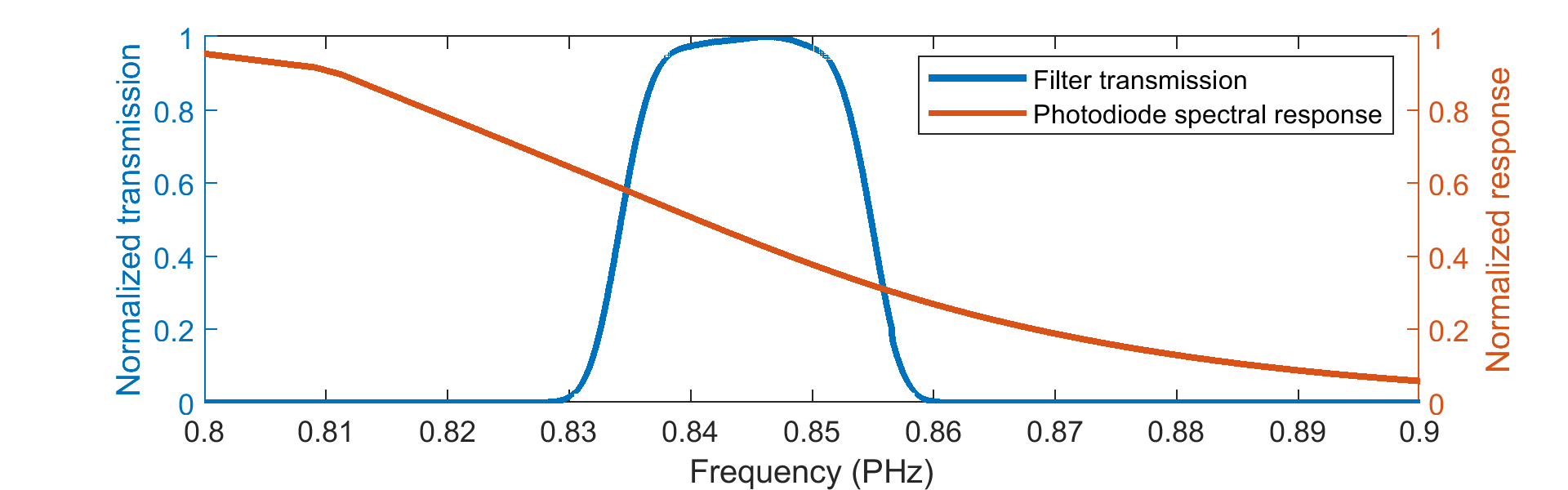}
	\caption{The spectral response of the photo-diode and the measured transmission of the bandpass filter used in experiment.}
	\label{diode_response}
\end{figure}

To get a more accurate experimental spectral response of detection, as well as to perform simulations close to experimental conditions, we took into account the spectral sensitivity of the photodiode (ALPHALAS GmBH) and the transmission of the bandpass filter (Thorlabs) (Fig.~\ref{diode_response}).
 ALPHALAS GmBH provided the spectral sensitivity of the photodiode.

\subsection{Influence of propagation effects and spectral response formation}
\label{bandpass_dependence}

\vskip 4mm

When sampling and test pulses have linear polarizations and are orthogonal to each other, polarizations of the produced local oscillator and heterodyne signal will depend on the symmetry of a nonlinear medium.
For instance, a $z$-cut $\alpha$-quartz crystal has both $\chi^{(2)}$ and $\chi^{(3)}$ nonlinearities.
Although the $\chi^{(3)}$ nonlinearity is independent on rotating the crystal around its axis, the $\chi^{(2)}$ tensor allows one to control both the magnitude and direction of the second-order nonlinear polarization.
This can be done by a simple rotation of the crystal.
In particular, there is a rotation angle that makes the second harmonic of the sampling pulse to be polarized orthogonally to the sampling pulse.
The same rotation angle makes the sum- and difference-frequency generation between the orthogonally polarized sampling and test pulses generate a nonlinear polarization that has a nonzero component along the electric field of the sampling pulse.
In this case, a polarizer that transmits the sampling pulse and blocks the test pulse will make the SHG/DFG+THG to be the dominant GHOST mechanism.
By keeping the orientation of  $z$-cut $\alpha$-quartz crystal and rotating the polarizer into a position where it transmits the test pulse and blocks the sampling pulse, one will select the FWM/XPM+SHG GHOST.
It is because, in this case, the $\chi^{(3)}$ nonlinearity that mixes one photon of the test pulse with two photons from the sampling pulse generates a polarization response that is parallel to the test pulse.
Thus, the polarizer placed  after the nonlinear medium allows one to choose between various GHOSTs.

\begin{figure}[ht!]
\centering
	\includegraphics[scale=0.3]{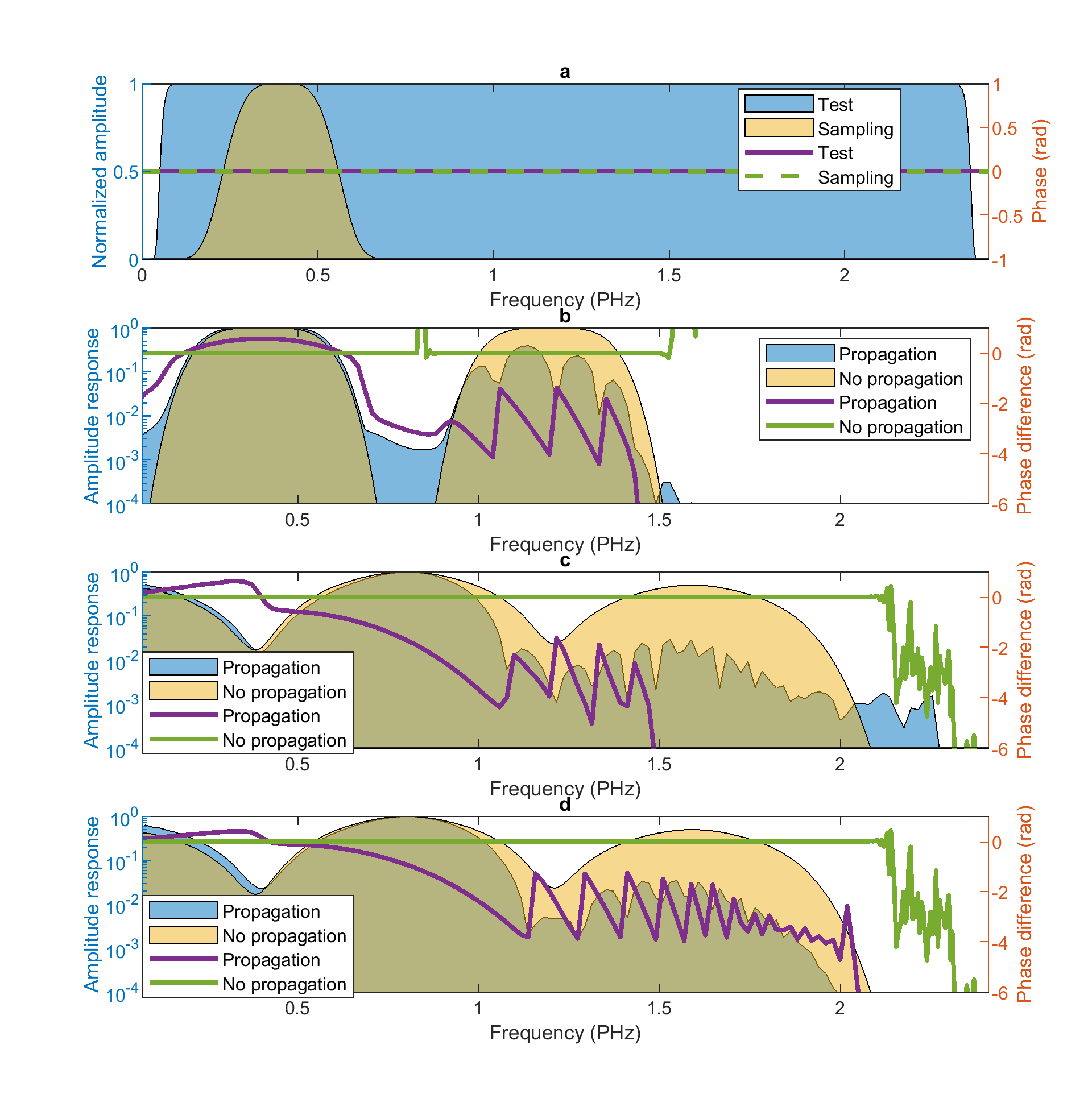}
	\caption{(\textbf{a}) Fourier representation of orthogonally polarized sampling and test pulses used for simulations.
	(\textbf{b}) and (\textbf{c}) are simulated spectral responses of a 10~{\textmu}m thick $z$-cut $\alpha$-quartz crystal oriented such that a second harmonic from the sampling pulse is orthogonal to a sampling polarization.
	(\textbf{b}) corresponds to a detection along the polarization of a sampling pulse, while (\textbf{c}) corresponds to the detection along the polarization of the test pulse.
	(\textbf{d}) simulated spectral response of 10~{\textmu}m thick fused silica medium, followed by a polarizer at $45^\circ$ with respect to sampling and test polarizations.
	Filled areas depict normalized spectral amplitude responses, while solid lines depict spectral phase differences.
	The detection frequency was set to 0.8~PHz (center of a second harmonic of the sampling pulse).}
	\label{prop1}
\end{figure}

Figure \ref{prop1} shows a typical spectral response simulated with and without propagation though a 10~{\textmu}m thick  $z$-cut $\alpha$-quartz crystal and 10~{\textmu}m thick fused silica glass.
The simulation is based on solving Maxwell's equations as described section \ref{propagation_model}.
The response without propagation, obtained under the assumption of an instantenous nonlinear polarization response, is shown for comparison.

\subsection{Spectral responses for NPS and GHOST measurements}

\vskip 4mm

\begin{figure}[ht!]
\centering
	\includegraphics[scale=0.3]{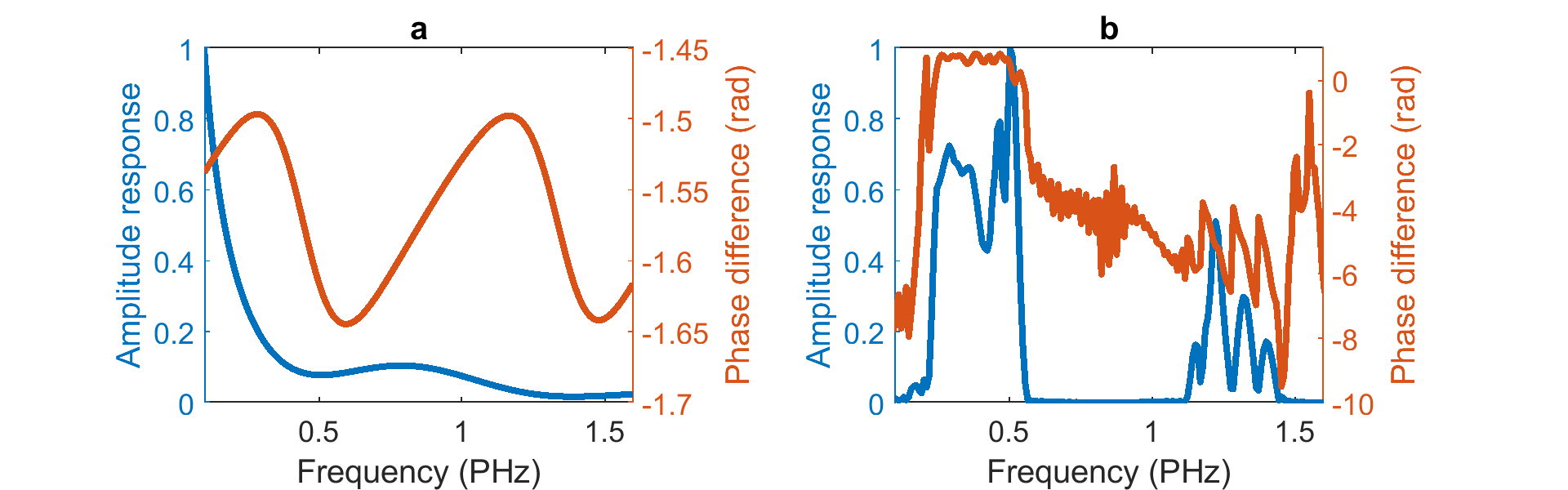}
	\caption{Calculated spectral responses of NPS (\textbf{a}) and GHOST (SFG+THG and DFG+THG) (\textbf{b}) techniques, based on the measured sampling pulse.} 
	\label{spectral_responses}
\end{figure}

In order to confirm the detection of a test pulse waveform based on SFG+THG and GHOST, the same test pulse waveform was measured with GHOST and NPS techniques.
To reconstruct the electric field of the test pulse, the spectral responses of NPS and GHOST techniques were determined.
For this purpose, the sampling pulse was characterized with a calibrated grating spectrometer and with a PG FROG technique.
From the characterized sampling pulse, the spectral response of the NPS technique can be easily obtained as described in the original publication\cite{Sederberg2020}.
To obtain the GHOST spectral response, the propagation of a characterized sampling pulse together with a known broadband theoretical pulse was simulated.
Fig.~\ref{spectral_responses} shows the obtained experimental responses of both techniques for experimental sampling pulse with CEP 0.
In the experiment presented in the main text, the CEP of the sampling pulse was set to 0 with a solid-state light-phase detection technique~\cite{Paasch-Colberg2014}.

\subsection{Field strength and CEP dependence of GHOST channels}

\vskip 4mm

Different GHOSTs scale differently with intensity of the sampling pulse, different spectral detection limits, and have a different CEP offsets, which are always an integer multiple of the CEP of the sampling pulse.
Table \ref{tab_1} summarizes several GHOSTs based on SHG and THG local oscillators.
The bandwidth limits are established for the case of a rectangular spectrum of the sampling pulse, having a constant amplitude in the range from $\omega-\Delta_{\omega}/2$ to $\omega + \Delta_{\omega}/2$ and zero otherwise.
The highest and lowest detectable frequencies are then given by the signals overlapping with the extrema of the up-converted local-oscillator band.
The bandwidth values tabulated below are those values at which such a signal is not strictly zero.
For a real pulse, the shape of the spectral response will depend on both its spectral amplitudes and spectral phases, as well as the phase-matching conditions of the nonlinear medium, filtration, and photodetector response. 

\begin{table}[ht!]
\begin{center}
\begin{tabular}{  c | c | c | c | c | c } 
 LO & HS & Scaling & Lower BW limit & Higher BW limit & $\Delta \phi_\mathrm{CE}$ \\ 
 \hline
 SHG & SFG & $E_\mathrm{T}E_\mathrm{S}^{3}$ & $\omega - \frac{3}{2}\Delta_{\omega}$& $\omega + \frac{3}{2}\Delta_{\omega}$ & $-\phi_\mathrm{S}$ \\ 
 \hline
 SHG & DFG & $E_\mathrm{T}E_\mathrm{S}^{3}$ & $3\omega - \frac{3}{2}\Delta_{\omega}$ & $3\omega + \frac{3}{2}\Delta_{\omega}$ & $\phi_\mathrm{S}$ \\ 
 \hline
 THG & SFG & $E_\mathrm{T}E_\mathrm{S}^{4}$ & $2\omega - 2\Delta_{\omega}$ & $2\omega + 2\Delta_{\omega}$ & $-2\phi_\mathrm{S}$ \\ 
 \hline
 THG & DFG & $E_\mathrm{T}E_\mathrm{S}^{4}$ & $4\omega - 2\Delta_{\omega}$ & $4\omega + 2\Delta_{\omega}$ & $4\phi_\mathrm{S}$ \\ 
 \hline
 SHG & FWM+, FWM- & $E_\mathrm{T}E_\mathrm{S}^{4}$ & 0  & $2\Delta_{\omega}$ & 0 \\ 
 \hline
 THG & FWM+ & $E_\mathrm{T}E_\mathrm{S}^{5}$ & $\omega - \frac{5}{2}\Delta_{\omega}$ & $\omega+\frac{5}{2}\Delta_{\omega}$ & $-\phi_\mathrm{S}$ \\ 
 \hline
 THG & FWM- & $E_\mathrm{T}E_\mathrm{S}^{5}$ & $5\omega - \frac{5}{2}\Delta_{\omega}$ & $5\omega+\frac{5}{2}\Delta_{\omega}$ & $5 \phi_\mathrm{S}$ \\ 
 \hline
 THG & XPM & $E_\mathrm{T}E_\mathrm{S}^{5}$ & $3\omega - \frac{5}{2}\Delta_{\omega}$ & $3\omega+\frac{5}{2}\Delta_{\omega}$ & $-3 \phi_\mathrm{S}$ \\ 
\end{tabular}
\end{center} 
\caption{GHOSTs based on second and third harmonic of the sampling pulse as the local oscillator.
$\omega$ is the central angular frequency of the sampling pulse.
$\Delta_{\omega}$ is the bandwidth of the sampling pulse.
$\Delta \phi_\mathrm{CE}$ is the shift of the CEP of the measured waveform relative to the actual electric field, which depends on the CEP of the sampling pulse, $\phi_\mathrm{S}$.}
\label{tab_1}
\end{table}

\subsection{Phase dependence of waveform detection}
\label{unstable_cep_section}

\vskip 4mm

The appearance of a stable waveform when measuring unstable electric fields can be understood through a detailed look at the measurement process.
It emerges that the CEP of the test waveform, and the fluctuations it contains, is canceled in the measurement through an equal and opposite contribution of the CEP of the sampling pulse.
The quantity that remains is a stable waveform containing similar information to what one would obtain from a traditional pulse measurement technique but measured in a way that allows sensitivity enhancement methods from field sampling and without further post-processing with an iterative algorithm.

The relationship between the measured signal in the case of the SFG+SHG GHOST and the actual electric field can be described as follows.
The SHG and SFG fields, $\tilde{E}_\mathrm{SHG}$ and $\tilde{E}_\mathrm{SFG}$, are either derived from the square of the sampling pulse, or the product of the sampling pulse with the test field.
In the frequency domain, these products become the convolution between the spectra:
\begin{equation}
    \tilde{E}_\mathrm{SHG}(\omega,\phi_\mathrm{S}) = \tilde{C}_\mathrm{SHG}(\omega)\int_{-\infty}^{\infty}d\omega'\,\tilde{E}_\mathrm{S}(\omega)e^{i\phi_\mathrm{S}}\tilde{E}_\mathrm{S}(\omega-\omega')e^{i\phi_\mathrm{S}} = e^{2 i \phi_\mathrm{S}}\tilde{E}_\mathrm{SHG}(\omega,0),
\end{equation}
\begin{equation}
    \tilde{E}_\mathrm{SFG}(\omega,\phi_\mathrm{S},\phi_\mathrm{T}) = \tilde{C}_\mathrm{SFG}(\omega)\int_{-\infty}^{\infty}d\omega'\,\tilde{E}_\mathrm{T}(\omega)e^{i\phi_\mathrm{T}}\tilde{E}_\mathrm{S}(\omega-\omega')e^{i\phi_\mathrm{S}} = e^{i (\phi_\mathrm{S} + \phi_\mathrm{T})}\tilde{E}_\mathrm{SFG}(\omega,0,0),
\end{equation}
\noindent where the CEPs of the sampling and test fields, $\phi_\mathrm{S}$ and $\phi_\mathrm{T}$, have been written explicitly, and the complex-valued constants $\tilde{C}_\mathrm{SHG}(\omega)$ and $\tilde{C}_\mathrm{SFG}(\omega)$ are determined by the nonlinear coefficients and phase-matching properties of the nonlinear medium, as well as their subsequent filtration and focusing onto the detector.

Their combined intensity on the detector is
\begin{multline}
    I_\mathrm{d}(\omega,\phi_\mathrm{S},\phi_\mathrm{T}) \propto \left| \tilde{E}_\mathrm{SHG}(\omega,\phi_\mathrm{S}) + \tilde{E}_\mathrm{SFG}(\omega,\phi_\mathrm{S},\phi_\mathrm{T})\right|^2\\=
     \left| \tilde{E}_\mathrm{SHG}(\omega,\phi_\mathrm{S})\right|^2 + \left| \tilde{E}_\mathrm{SFG}(\omega,\phi_\mathrm{S},\phi_\mathrm{T})\right|^2 + \left| \tilde{E}_\mathrm{SHG}(\omega,\phi_\mathrm{S}) \tilde{E}_\mathrm{SHG}(\omega,\phi_\mathrm{S})\right| \cos \left[\phi_\mathrm{T} - \phi_\mathrm{S} + \Phi_\mathrm{SFG}(\omega) - \Phi_\mathrm{SHG}(\omega)\right],
    \label{heterodyneF}
\end{multline}
where $\Phi_\mathrm{SHG}(\omega)$ and $\Phi_\mathrm{SFG}(\omega)$ are the spectral phases of $\tilde{E}_\mathrm{SHG}(\omega,0)$ and $\tilde{E}_\mathrm{SFG}(\omega,0,0)$, respectively.
The first two terms on the right-hand side are independent of $\phi_\mathrm{S}$ and $\phi_\mathrm{T}$, and only the final term, the cross term involving both fields, is relevant for field sampling.

In a typical measurement, one would introduce a time delay to one of the fields, introducing a linear phase $\omega\tau$ in the frequency domain.
The resulting modulation of the cross-term on the RHS of Eq.~\eqref{heterodyneF} then would trace out the measured waveform as the time delay $\tau$ was varied. 

Now, we can see where the CEPs of the input fields enter the measurement.
As mentioned previously, the SFG+SHG GHOST experiences a CEP shift of $-\phi_\mathrm{S}$, due to the appearance of this phase in Eq.~\eqref{heterodyneF}, unlike in, e.\,g., EOS, where all $\phi_\mathrm{S}$ contributions cancel.
Accordingly, $\phi_\mathrm{S}$ must be known and stabilized in order to determine the waveform of the test field.

We also note that in the case where $\phi_\mathrm{T}$ is fixed (for example, if the test field was derived from difference frequency generation), introducing a linear time dependence to $\phi_\mathrm{S}$ results in a sinusoidal variation of the cross term in Eq.~\eqref{heterodyneF}.
The time-averaged current will be zero, but a detectable high-frequency electronic signal will be introduced.
Thus, by introducing an offset of the frequency comb of the sampling pulse, an arbitrarily fast (up to the Nyquist frequency) modulation of this signal can be introduced, enabling sensitive detection when applied to MHz-repetition-rate laser systems with a controlled carrier-envelope offset frequency.

\bibliography{supplement} 
\bibliographystyle{naturemag-doi}